| | |
|---|---|
| Abstract | An atmospheric plasma torch has been supplied with a gaseous mixture of helium, water vapour and/or oxygen to study the production of reactive species within its flowing post-discharge, in particular hydrogen peroxide. The mechanisms responsible for the production of $H_2O_2$ have been investigated by correlating measurements of mass spectrometry and absorption photospectrometry. An absolute quantification of $H_2O_2$ has also been achieved and indicated that for 5 mmol of water vapour injected in the He-$H_2O$ discharge, approximately 9.5 µmol of $H_2O_2$ were produced in post-discharge. |

# 1. Introduction

The detection of hydrogen peroxide is an important issue in chemical, environmental and medical applications [1]. In sterilization, its toxic effect on bacteria can be enhanced by low concentrations of NO radicals [2] while in plasma medicine, $H_2O_2$ can either stimulate or inhibit cell proliferation or even induce cell apoptosis depending on its concentration [3], [4]. For these reasons, tailoring its concentration with high accuracy is crucial. This experimental work is focused on the mechanisms responsible for the production of $H_2O_2$ in a flowing post-discharge generated by an atmospheric plasma torch supplied in helium with/without water vapour and oxygen. Then, an absolute quantification of $H_2O_2$ detected in the flowing post-discharge has been achieved. According to the literature, manifold techniques have already been employed to measure $H_2O_2$ concentrations based on fluorescence, chemiluminescence, electrochemistry, titration and absorption spectrophotometry [5], [6]. Here, the later method has been chosen due to its high accuracy and its easy implementation to our plasma torch; it has been completed by mass spectrometry measurements.

# 2. Experimental setup

## 2.1. Plasma source

The plasma source is an RF atmospheric plasma torch from Surfx Technologies (Atomflo™ 400L-Series) supplied in helium (carrier gas), with the ability to mix reactive gases, i.e. water vapour and/or oxygen. The helium flow rate has been fixed to 20 L/min while oxygen and water vapour have been added for flow rates comprised between 0 and 150 mL/min. As sketched in Fig. 1, the resulting gas mixture (i) enters through the upper nozzle of the plasma torch, (ii) is uniformized through two perforated sheets, (iii) flows down around the left and right edges of the upper electrode (RF voltage) and (iv) passes through a slit at the center of the lower electrode (ground).

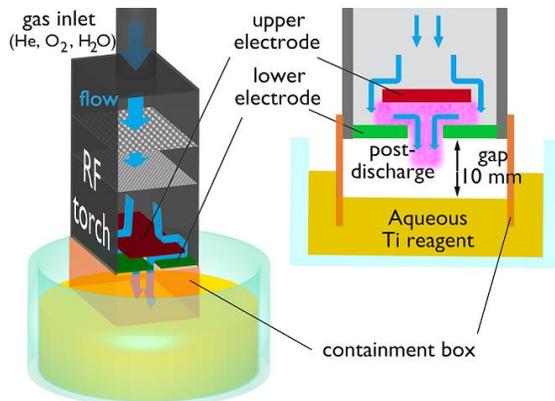

*Fig. 1. Scheme of the experimental setup.*

A reagent such as Fe (II), Co (II) or Ti (IV) compound (e.g. titanium oxysufate, titanium chlorate) is required for the conversion of hydrogen peroxide into a species photometrically detectable [7]. Therefore, a beaker containing an aqueous Ti reagent has been placed downstream the post-discharge so that a gap of 10 mm is maintained between the grounded electrode and the interface of the aqueous Ti reagent. As sketched in Fig. 1, a copper containment box confines the reactive species from the post-discharge to enhance their direct interaction with the reagent without any loss to the atmosphere.

## 2.2. Aqueous Ti reagent

An aqueous Ti reagent solution at a concentration of 320 mg/L has been realized by diluting 320 mg of Titanium (IV) oxide sulphate sulphuric acid hydrate powder ($TiOSO_4 \cdot xH_2O + H_2SO_4$) in sulphuric acid (20 mL, 16 M) completed by milli-Q water to obtain 1L (see reaction R1). As shown in reaction R2, this acidic aqueous solution of titanium oxysulfate reacts in presence of $H_2O_2$, leading to the formation of a yellow peroxotitanium complex $[Ti(O_2)OH(H_2O)_3]^+_{aq}$ whose absorbance (A) can be measured at 409 nm [8]. Then the hydrogen peroxide concentrations can be calculated using the Beer's law, provided $0.2 \leq A \leq 0.8$.

$$TiOSO_{4(s)} + 5H_2O \leftrightarrow [Ti(OH)_3(H_2O)_3]^+_{(aq)} + HSO^-_{4(aq)} \quad R1$$

$$[Ti(OH)_3(H_2O)_3]^+_{(aq)} + H_2O_{2(aq)} \leftrightarrow [Ti(O_2)(OH) \times (H_2O)_3]^+_{(aq)} + 2H_2O \quad R2$$

## 2.3. Mass spectrometry

Mass spectrometry measurements have been performed with a Hyden analytical QGA mass spectrometer using an ionization energy set to 35 eV. During these experiments, neither liquid reagent nor containment box were used; the capillary tube has been placed 10 mm downstream from the head of the plasma torch and maintained parallel to the post-discharge flow.

## 3. Results & discussion

### 3.1. Influence of the $O_2$ flow rate

An increasing flow rate of $O_2$ comprised between 0 and 150 mL/min has been added to the He flow rate (20 L/min) without any injection of water vapour. First, the species generated by the post-discharge have been evidenced by mass spectrometry (see Fig. 2). The peak intensities of He, $H_2O$ and OH remain constant since (i) the He flow rate is unchanged whatever the $O_2$ flow rate and (ii) the water vapour is permanently provided by the ambient atmosphere. Simultaneously, atomic and molecular oxygen show clear increasing peak intensities: approximately $4.10^{-5}$-$8.10^{-5}$ a.u. for $O_2$, $1.10^{-6}$-$6.10^{-6}$ a.u. for O radicals and $10^{-9}$-$10^{-8}$ a.u. for $O_3$. It is important to stress that no $H_2O_2$ species has been detected, whatever the $O_2$ flow rate is.

However, as presented in Fig. 3, the aqueous Ti reagent shows a slight absorbance (A≈0.1) at 409 nm regardless of the $O_2$ flow rate. As A<0.2 the Beer's law do not allow deducing the corresponding concentration of $H_2O_2$ with high accuracy but at least it evidences a non-negligible amount of $H_2O_2$ molecules in the reagent since a peak is observed. As A is almost constant for $\Phi(O_2)$=0-150 mL/min, it may be correlated with species from the post-discharge presenting a similar trend. Only the OH radicals satisfy this condition. Then, a recombination of OH radicals into $H_2O_2$ molecules could occur at the gas/liquid interface and explain why they are undetected in the gaseous phase but detected in the aqueous Ti reagent.

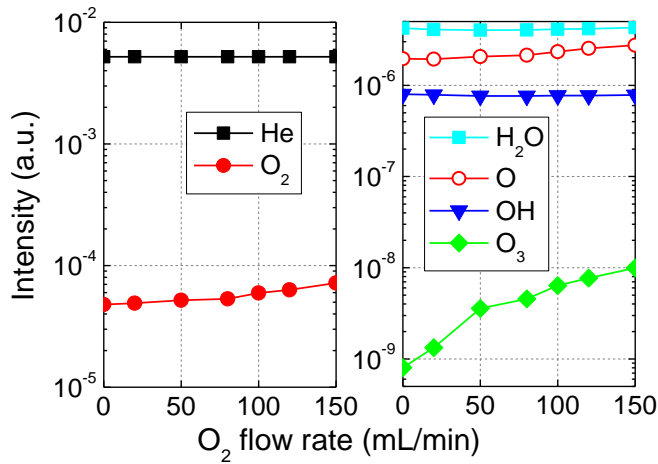

*Fig. 2. Peak intensities of species detected in the flowing post-discharge for $\Phi(He)$=20 L/min, $\Phi(O_2)$=0 mL/min, $P_{RF}$=120W, gap=10 mm.*

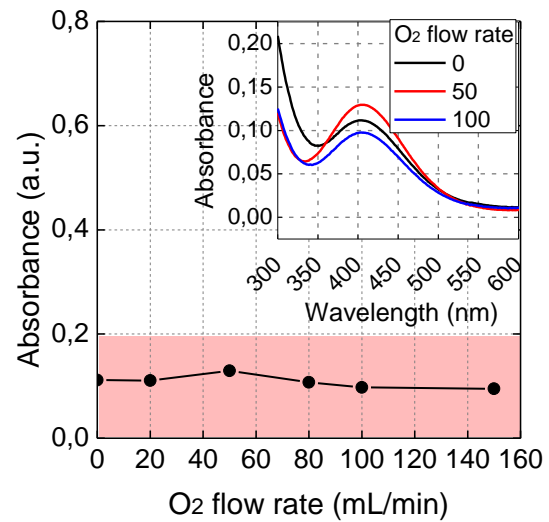

*Fig. 3. Absorbance of the aqueous Ti reagent after exposure to the flowing post-discharge for various $O_2$ flow rates (exp. time=5 min).*

### 3.2. Influence of the water vapour flow rate

Water vapour has been mixed with the carrier gas for flow rates increasing from 0 to 130 mL/min, without any admixture of $O_2$ gas. The mass spectrometry measurements reported in Fig. 4 show trends very different from those obtained in Fig. 2. Indeed, the intensities of O, $O_2$ and $O_3$ remain constant, while the concentration of OH strongly increases as well as the intensity of $H_2O$. Also, the intensity of $H_2O_2$ is now detected but remains low and constant (about $1.10^{-6}$ a.u.) on the entire flow rate range. These results have been correlated with absorption photospectroscopy measurements, plotted in Fig. 5. As the volume of the reagent is known (25 mL), applying the Beer's law allows evaluating the number of $H_2O_2$ moles satisfying the reaction R2, e.g. 0.5 µmol in a pure He post-discharge while 9.5 µmol in a He-$H_2O$ post-discharge with $\Phi(H_2O)$=130 mL/min. The production of $H_2O_2$ in the aqueous Ti reagent could mainly result from the combination of two OH radicals in the gaseous phase (OH + OH → $H_2O_2$) since $H_2O_2$ is known to be stable and its diffusion is a significant loss mechanism [9]. Surprisingly, the intensity of $H_2O_2$ seems constant in Fig. 4, instead of increasing with the intensity of the OH radicals. Two assumptions may explain this apparent plateau. First, an alternative pathway can be responsible for the permanent increase of $H_2O_2$ in the aqueous Ti reagent (see Fig. 5) while not in the gas. In particular, the existence of a liquid interface may favour the production of $H_2O_2$ molecules with OH radicals through 3 body-collisions. Second, this plateau corresponds to a saturation of the aforementioned reaction, in that the post-discharge is confined in the containment box and only a new amount of $H_2O_2$ molecules can be generated if the same amount of $H_2O_2$ molecules has reacted in the aqueous Ti reagent.

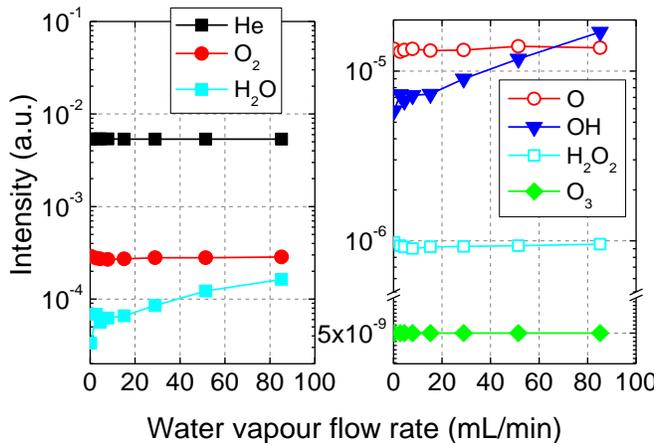

Fig. 4. Peak intensities of species detected in the flowing post-discharge for $\Phi(He)$=20 L/min, $\Phi(O_2)$=0 mL/min, $P_{RF}$=120W, gap=10 mm.

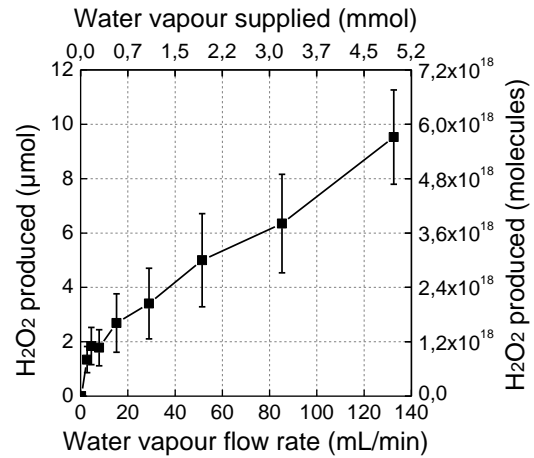

Fig. 5. Amount of hydrogen peroxide produced in the flowing post-discharge as a function of the water vapour flow rate (process time=1 min).

### 3.3. Influence of the $O_2$ flow rate in presence of water vapour

An increasing flow rate of $O_2$ comprised between 0 and 150 mL/min has been added to the He flow rate (20 L/min) for $\Phi(H_2O)$=30 mL/min. As expected, the intensity of He is unchanged while rises in the intensities of O, $O_2$ and $O_3$ are observed. Also, the intensity of OH is constant (consistently with its trend in Fig. 2) and the intensity of $H_2O$ as well since $\Phi(H_2O)$ is fixed. However – contrarily to the He-$H_2O$ discharge – the production of hydrogen peroxide has been evidenced, increasing from $9.7 \cdot 10^{-7}$ (pure helium post-discharge) to $1.5 \cdot 10^{-6}$ a.u (150 mL/min of $O_2$). This result is in agreement with the one-dimensional fluid simulations of K. McKay et al where the highest $H_2O_2$ densities (as high as $1.2 \cdot 10^{15}$ cm$^{-3}$) have been obtained for a helium discharge with 0.3% of water vapour and 1% of $O_2$ [9]. The combination of OH radicals can explain the production of $H_2O_2$ but an alternative channel could also be opened, i.e. the production of O radicals (from the dissociation of $O_2$) and their subsequent attachment to the excited $H_2O$ molecules.

The Fig. 7 shows the corresponding amounts of $H_2O_2$ produced in the post-discharge and detected by absorption photospectrometry. No clear trend has been draught but interesting is the average amount of $H_2O_2$ produced. For instance, at $\Phi(O_2)$=30 mL/min, almost 15 µmol of $H_2O_2$ have been produced in this He-$H_2O$-$O_2$ discharge against 3.3 µmol in the He-$H_2O$ discharge (see Fig. 5). As a consequence, the hydrogen peroxide production can be tailored either by rising the water vapour flow rate or by injecting some oxygen.

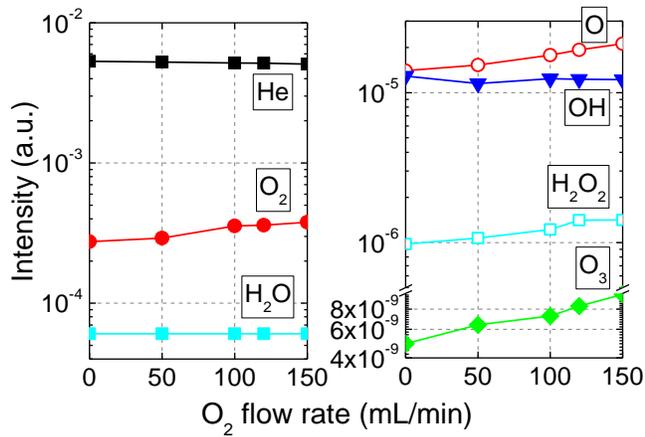

Fig. 6. Peak intensities of species detected in the flowing post-discharge for $\Phi(He)$=20 L/min, $\Phi(H_2O)$=30 mL/min, $P_{RF}$=120W, gap=10 mm.

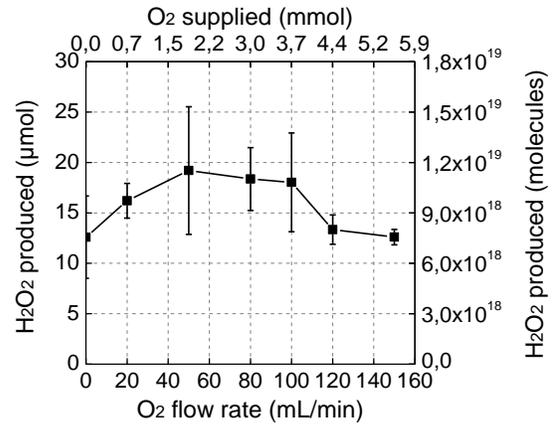

Fig. 7. Amount of hydrogen peroxide produced in the flowing post-discharge as a function of the water vapour flow rate (process time=1 min, $\Phi(H_2O)$=30 mL/min).

# 4. Conclusion

The production of hydrogen peroxide in an atmospheric flowing post-discharge has been demonstrated using water vapour as reactive gas. The $H_2O_2$ concentration can also be tailored with high accuracy either by tuning the water vapour flow rate or by mixing some oxygen to the He-$H_2O$ discharge (while maintaining the same total flow rate). The combination of two OH radicals in the post-discharge appears as the most reliant reaction to explain the production of $H_2O_2$, but alternative channels may occur as well and will be the subject of further investigations.

# 5. Acknowledgment

This work has been supported by the PSI-IAP 7 (Plasma Surface Interactions) from the Belgian Federal Government BELSPO agency).

# 6. References

[1] D. Dobrynin, G. Fridman, G. Friedman, A. Fridman, New Journal of Physics, **11** (2009), 115020.

[2] R. S. Dawe, Br. J. Dermatol., **149** (2003), 669-672.

[3] M. G. Kong, G. Kroesen, G. Morfill, T. Nosenko, T. Shimizu, J. van Dijk, J. L. Zimmermann, New Journal of Physics, **11** (2009), 115012

[4] V. J. Thannickal, B. L. Fanburg, Lung. Cell Mol. Physiol. **279** (2000) L1005-28.

[5] M. C. Ramos, M. C. Torijas, A. Navas Diaz, Sensors and Actuators B: Chemical, 2001, **73**(1), 71-75.

[6] N. V. Klassen, D. Marchington, H. C. E. McGowan, Anal. Chem. (1994), 66(18), 2921-2925

[7] A. Pashkova, K. Svajda, G. Black, R. Dittmeyer, Review of scientific instruments, 80, 055104 (2009)

[8] K. P. Reis, V. K. Joshi, M. E. Thompson, Journal of catalysis (1996), **161**, 62-67.

[9] K. McKay, D. X. Liu, M. Z. Rong, F. Iza, M. G. Kong, J. Phys. D: Appl. Phys., **45** (2012), 172001 (5pp).